\documentclass[11pt]{amsart}

\usepackage{epstopdf}
\usepackage{graphicx}
\usepackage{amssymb}
\usepackage{url}
\usepackage{mathrsfs}

	\normalbaselines						  %

\newcommand{\Z}{{\mathbb  Z}}

\newcommand{\fdot}{\,\cdot\,}

\makeatletter
\def\Ddots{\mathinner{\mkern1mu\raise\p@
\vbox{\kern7\p@\hbox{.}}\mkern2mu
\raise4\p@\hbox{.}\mkern2mu\raise7\p@\hbox{.}\mkern1mu}}
\makeatother

\catcode`@=11
\chardef\mathlig@atcode\count255

\def\actively#1#2{\begingroup\uccode`\~=`#2\relax\uppercase{\endgroup#1~}}
\def\mathlig@gobble{\afterassignment\mathlig@next@cmd\let\mathlig@next= }
\def\mathlig@delim{\mathlig@delim}
\def\mathlig@defcs#1{\expandafter\def\csname#1\endcsname}
\def\mathlig@let@cs#1#2{\expandafter\let\expandafter#1\csname#2\endcsname}
\def\mathlig@appendcs#1#2{\expandafter\edef\csname#1\endcsname{\csname#1\endcsname#2}}

\def\mathlig#1#2{\mathlig@checklig#1\mathlig@end\mathlig@defcs{mathlig@back@#1}{#2}\ignorespaces}

\def\mathlig@checklig#1#2\mathlig@end{%
 \expandafter\ifx\csname mathlig@forw@#1\endcsname\relax
 \expandafter\mathchardef\csname mathlig@back@#1\endcsname=\mathcode`#1%
 \mathcode`#1"8000\actively\def#1{\csname mathlig@look@#1\endcsname}%
 \mathlig@dolig#1\mathlig@delim
\fi
\mathlig@checksuffix#1#2\mathlig@end
}

\def\mathlig@checksuffix#1#2\mathlig@end{%
\ifx\mathlig@delim#2\mathlig@delim\relax\else\mathlig@checksuffix@{#1}#2\mathlig@end\fi
}
\def\mathlig@checksuffix@#1#2#3\mathlig@end{%
\expandafter\ifx\csname mathlig@forw@#1#2\endcsname\relax\mathlig@dosuffix{#1}{#2}\fi
\mathlig@checksuffix{#1#2}#3\mathlig@end
}

\def\mathlig@dosuffix#1#2{%
\mathlig@appendcs{mathlig@toks@#1}{#2}%
\mathlig@dolig{#1}{#2}\mathlig@delim
}

\def\mathlig@dolig#1#2\mathlig@delim{%
 \mathlig@defcs{mathlig@look@#1#2}{%
 \mathlig@let@cs\mathlig@next{mathlig@forw@#1#2}\futurelet\mathlig@next@tok\mathlig@next}%
 \mathlig@defcs{mathlig@forw@#1#2}{%
  \mathlig@let@cs\mathlig@next{mathlig@back@#1#2}%
  \mathlig@let@cs\checker{mathlig@chck@#1#2}%
  \mathlig@let@cs\mathligtoks{mathlig@toks@#1#2}%
  \expandafter\ifx\expandafter\mathlig@delim\mathligtoks\mathlig@delim\relax\else
  \expandafter\checker\mathligtoks\mathlig@delim\fi
  \mathlig@next
 }%
 \mathlig@defcs{mathlig@toks@#1#2}{}%
 \mathlig@defcs{mathlig@chck@#1#2}##1##2\mathlig@delim{%
  \ifx\mathlig@next@tok##1%
   \mathlig@let@cs\mathlig@next@cmd{mathlig@look@#1#2##1}\let\mathlig@next\mathlig@gobble
  \fi
  \ifx\mathlig@delim##2\mathlig@delim\relax\else
   \csname mathlig@chck@#1#2\endcsname##2\mathlig@delim
  \fi
 }%
 \ifx\mathlig@delim#2\mathlig@delim\else
  \mathlig@defcs{mathlig@back@#1#2}{\csname mathlig@back@#1\endcsname #2}%
 \fi
}%

\catcode`@\mathlig@atcode

\mathchardef\ordinarycolon\mathcode`\:
\def\vcentcolon{\mathrel{\mathop\ordinarycolon}}
\mathlig{:=}{\vcentcolon=}
\mathlig{::=}{\vcentcolon\vcentcolon=}

\numberwithin{equation}{section}

\theoremstyle{plain}

\theoremstyle{definition}

\theoremstyle{remark}

\theoremstyle{remark}
\newtheorem*{exs*}{Examples}
\theoremstyle{remark}
\newtheorem*{rems*}{Remarks}
\theoremstyle{remark}
\newtheorem*{rem*}{Remark}

\title[Conductance in two dimensional random media with small disorder]{Conductance for the two dimensional discrete random Schr\"odinger operator with small disorder}
\author{Constanze~Liaw}
\thanks{The author is partially supported by the NSF grant DMS-1101477.}
\address{Department of Mathematics, Baylor University, One Bear Place $\#$97328, Waco, TX  76798-7328, USA}
\email{Constanze$\underline{\,\,\,\,}$Liaw@baylor.edu}

\begin{document}

\begin{abstract}
\medskip\noindent As part of condensed-matter physics, the field of Anderson localization concerns the study of conductance of electrons in a random medium.

We consider the discrete random Schr\"odinger operator on the integer lattice $\Z^2$. Based on a recent mathematical result we introduce a new numerical approach to the Anderson (de)localization problem. As an application we show numerically the following unexpected result: For small disorder, this random operator allows extended states with positive probability. This approach eliminates potential problems due to boundary effects, and the numerical part is rather simple compared with other experiments in the field. Further, we provide a mathematical derivation of a quantity closely related to Thouless' dimensionless scaling parameter.
No new information is gained regarding the energy regimes at which diffusion occurs.
\end{abstract}

\maketitle

\section{Introduction}

In 1958 P.W.~Anderson \cite{And1958} suggested that sufficiently large impurities in a semi-conductor could lead to spatial localization of electrons, called Anderson localization.
The field has grown into a rich theory and is studied by both, the physics and the mathematics community.

We consider the discrete random Schr\"odinger operator on the integer lattice of dimension two $H_\omega = - \bigtriangleup + \sum_{i\in \Z^2} \omega_i <\fdot, \delta_i> \delta_i$ and assume that the random variables $\omega_i$ are independent identically distributed with uniform distribution in $[-c,c]$, i.e.~each number in the interval is attained with equal probability. This operator describes the situation where the atoms of the crystal are located `near' the integer lattice points $\Z^2$.

Many approaches to Anderson localization are based on scaling theory. A dimensionless scaling parameter \cite{1972article} is the sole indicator whether or not the system exhibits Anderson localization. The parameter $g$ is the ratio between the Heisenberg time (the maximum time that a wave packet can travel inside a finite region before revisiting the same location) and the Thouless time (the time it takes an extended state to arrive at the boundary of a finite region). Roughly speaking, for $g<1$ we have localization, and otherwise the existence of extended states.
Our results are in contradiction with much numerical and experimental work using the scaling theory (see e.g.~\cite{brandeskettenmann2003, kramer1993, physicstoday2009}) which indicates that one should expect some ``marginal localization'' in the form of so-called quasi-extended states in dimension two for small disorder.

However, mathematically rigorous proofs for the dynamical localization of the discrete random Schr\"odinger operator on $l^2(\Z^d)$ are only known in dimension one at all strengths \cite{CL, CFKS, FP} and in dimension two and higher for disorders $c$ above a certain threshold \cite{AizMol1993, FS, SIMREV}.
For the discrete random Schr\"odinger operator on $l^2(\Z^d)$ there is no proof, whether or not diffusion occurs in dimension two and higher for small disorder $c$. Many results are known for related operators, e.g.~the discrete random Schr\"odinger operator on the tree and other structures.

Based on a recent mathematical result \cite{AbaLiawPolt} -- which is based on the study of rank one perturbations -- the author designed a new numerical approach to indicate conductance in random media. The key to making this method numerically feasible is that it suffices to track the evolution of just one vector under the repeated application of the random operator. The details of this numerical experiment, the mathematical background, and related information can be found in \cite{Exp}.

In Section \ref{s-setup}, we explain the idea behind the experiment and how it indicates that the discrete random Schr\"odinger operator in two dimensions does in fact exhibit diffusion for small disorders $c \lesssim 0.7$ with positive probability.
Via variations of the code we study the evolution over time of the energy of a wave packet initially located at the origin, as well as the dependency with respect to $c$ of the location of the energy of such a wave packet in Section \ref{s-scaling}. We show how those mathematical and numerical results are related to the Thouless parameter, and conclude the existence of extended states in dimension two for small disorder.

The method within can be applied to many operators from a larger class of discrete Anderson models, so-called Anderson-type Hamiltonians \cite{JakLast2000}. The author expects a similar result for the discrete random Schr\"odinger operator in dimension three and higher but has not investigated those cases. However, this method cannot not (yet) be applied to two other important random Hamiltonians: Due to memory restrictions, the method is not suited well for the discrete random Schr\"odinger operator on the tree. And the mathematical result cannot be applied for the Rademacher potential where the random variables attain the values $\pm 1$ with equal probability.

\section{Key steps of the new approach}\label{s-setup}
Consider the discrete random Schr\"odinger operator $H_\omega$ on $l^2(\Z^2)$ with identically distributed random variables $\omega_{ij}$ with uniform distribution in $[-c,c]$. Fix the vectors $\delta_{00}\in l^2(\Z^2)$ and $\delta_{11}\in l^2(\Z^2)$, namely
\begin{align}\label{e-example}
\delta_{00}=
\begin{pmatrix}
\ddots&\vdots&\vdots&\vdots&\vdots&\vdots&\Ddots\\
\hdots&0&0&0&0&0&\hdots\\
\hdots&0&0&0&0&0&\hdots\\
\hdots&0&0&1&0&0&\hdots\\
\hdots&0&0&0&0&0&\hdots\\
\hdots&0&0&0&0&0&\hdots\\
\Ddots&\vdots&\vdots&\vdots&\vdots&\vdots&\ddots
\end{pmatrix}\,,\qquad\quad
\delta_{11}=
\begin{pmatrix}
\ddots&\vdots&\vdots&\vdots&\vdots&\vdots&\Ddots\\
\hdots&0&0&0&0&0&\hdots\\
\hdots&0&0&0&0&0&\hdots\\
\hdots&0&0&0&0&0&\hdots\\
\hdots&0&0&0&1&0&\hdots\\
\hdots&0&0&0&0&0&\hdots\\
\Ddots&\vdots&\vdots&\vdots&\vdots&\vdots&\ddots
\end{pmatrix} \,.
\end{align}

Consider the distance $D_{\omega,c}^n$ between the unit vector $\delta_{11}$ and the subspace obtained by taking the span of the vectors $\{\delta_{00}, H_\omega \delta_{00}, H_\omega^2 \delta_{00}, \hdots, H_\omega^n \delta_{00}\}$. In other words, let
\begin{align*}
D_{\omega,c}^n := \text{dist}(\delta_{11}, \text{span}\{H_\omega^k \delta_{00}:k=0,1,2,\hdots,n\}).
\end{align*}

Deep mathematical results (see Corollary 3.3 of \cite{Exp}) imply that dynamical delocalization takes place, if we can find a disorder $c>0$ for which the distance $D_{\omega,c}^n$ does not tend to zero as $n$ approaches $\infty$, i.e.~$D_{\omega,c}>0$ where
$$
D_{\omega,c} := \lim_{n\to\infty} D_{\omega,c}^n.
$$

It is worth mentioning that this mathematical result is a `one-sided' implication. In particular, if $D_{\omega,c} = 0$, then we cannot conclude localization.

Some fairly simple arithmetic shows that $D_{\omega,c}$ is given by the following expression
\begin{align}\label{e-dinfty}
D_{\omega,c} = \sqrt{1 - \sum_{k=0}^{\infty} \frac{<m_k, \delta_{11}>^2}{\|m_k\|_2^2}}\,,
\end{align}
where $\{m_0, m_1, \hdots, m_n\}$ form the orthogonal basis of the linear subspace spanned by $H_\omega^k \delta_{00}$ for $k=0,1,2,\hdots,n$ of $l^2(\Z^2)$ obtained by the Gram--Schmidt process (without normalization), and $\|\cdot\|_2$ denotes the Euclidean norm.

In the numerical experiments, in order to exclude the possibility that the distance $D_{\omega,c}^n$ tends to zero logarithmically, we re-scaled time and applying the negative exponent to the horizontal axis which satisfied the least squares property for line approximation. The intersection $y_{\omega,c}$ of the approximating line with the vertical axis is an estimate for the distance $D_{\omega,c}$.
Further we considered the lower estimate $L_{\omega,c}$ for $y_{\omega,c}$ obtained by taking the lowest $y-$intercept lines through any two consecutive points.
We repeated the experiment for many realizations of the random variable $\omega$ and chose the smallest values of $y_{\omega,c}$ and $L_{\omega,c}$ obtained from all randomizations. Further, we considered a wide range of disorders $c$.

\begin{figure}
 \centerline{
 \includegraphics[width=4.7in]{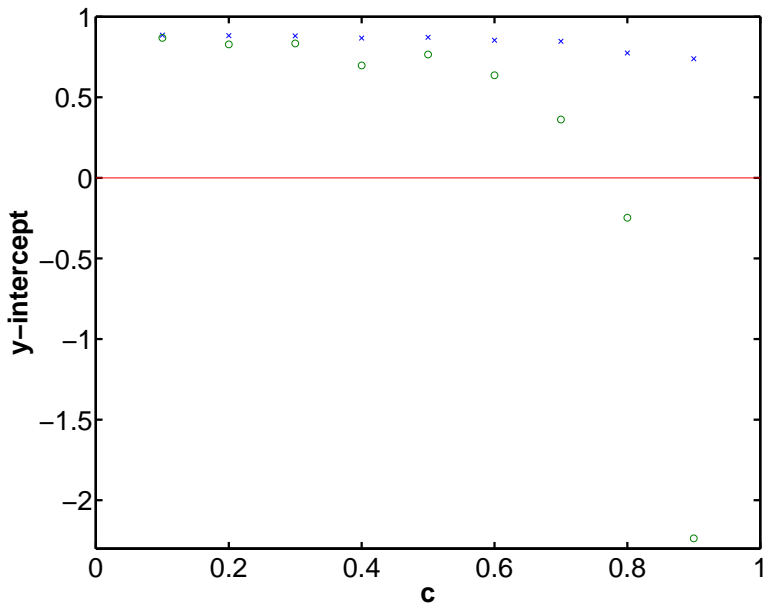}
 }
\caption{As a function of $c$ we show $y_{\omega,c}$ ($x$'s; larger function values) and $L_{\omega,c}$ (circles; smaller function values). We conclude $D_{\omega,c} \approx y_{\omega,c} \ge L_{\omega,c}>0$ for $c\lesssim0.7$.}\label{mainfigure}
\end{figure}

We include Figure \ref{mainfigure} (see \cite{Exp}) which shows these smallest values of $y_{\omega,c}$ and $L_{\omega,c}$ as a function of disorder $c$.

\section{Connection with the scaling parameter and existence of extended states}\label{s-scaling}
We explain how the delocalization result described in Section \ref{s-setup} implies the existence of extended states in the sense of the Thouless criterion.

The numerator $<m_k, \delta_{11}>$ in formula \eqref{e-dinfty} is precisely the $(1,1)-$entry of the vector $m_k$, while the denominator is the total energy of $H_\omega^k\delta_{00}$ after successive orthogonalization (we only consider the new information of each state).

Roughly speaking, the numerator is small often precisely if the Heisenberg time is large, i.e.~it takes a long time to return to the same finite region. And if much of the energy of the state $m_k$ is located away from the origin, then the Thouless time is small.

\begin{figure}
 \centerline{
 \includegraphics[width=4.9in]{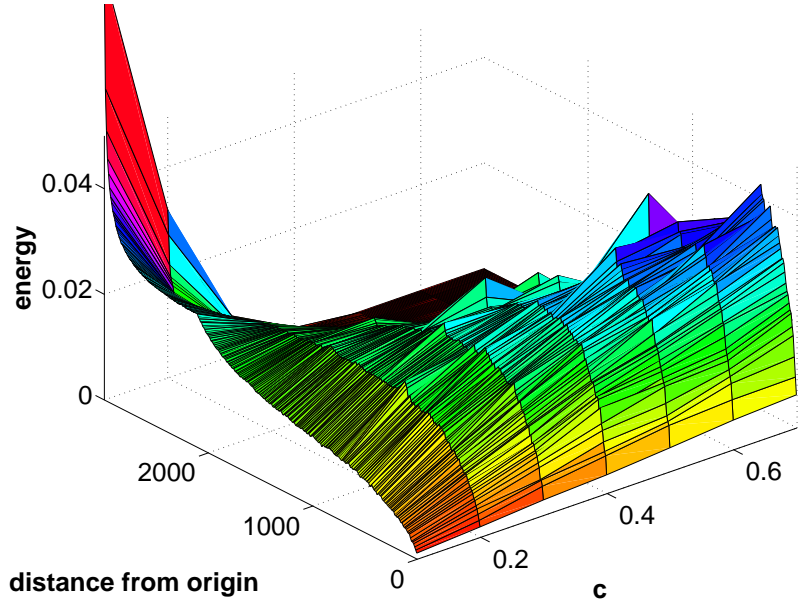}
 }
\caption{The figure shows the energy distribution of $m_{3000}$ as a function of distance from the origin for several values of disorder $c$.}\label{f-energy2}
\end{figure}

In the case of small $c$, Figure \ref{f-energy2} (see \cite{Exp}) shows that most of the energy of $m_k$ is located `away' from the origin almost surely. In particular, $<m_k, \delta_{11}>$ is small often and $m_k$ is relatively large for all $k$. Indeed, a large part of the energy is almost surely located on the outermost diamond which the orthogonalization process leaves unchanged. In particular, the norm $\|m_k\|$ remains comparably large. This implies that the Heisenberg time is large and the Thouless time is small, yielding a scaling parameter $>1$ and the existence of extended states (in the physicists' sense), i.e.~conductance of electrons. As was mentioned above, our method of computing a lower estimate of $D_{\omega,c}$ also indicates dynamical delocalization and hence the existence of extended states (in the mathematical sense meaning states that do not decay exponentially, and even absolutely continuous spectrum).

As the disorder increases, Figure \ref{f-energy2} shows that a larger part of the energy remains near the origin, so that $<m_k, \delta_{11}>$ is away from zero more often and the ratio $\frac{<m_k, \delta_{11}>^2}{\|m_k\|_2^2}$ is generally larger than for small $c$.
However, in this case a large part of the energy remains near the origin. The Euclidean norm $\|m_k\|_2$ in the denominator of \eqref{e-dinfty} has little to do with the Thouless time. Therefore, the relationship between the ratio in formula \eqref{e-dinfty} and the scaling parameter is not directly obvious. This is in agreement with the fact that our new method cannot be used to show localization.

\providecommand{\bysame}{\leavevmode\hbox to3em{\hrulefill}\thinspace}
\providecommand{\MR}{\relax\ifhmode\unskip\space\fi MR }
\providecommand{\MRhref}[2]{%
  \href{http://www.ams.org/mathscinet-getitem?mr=#1}{#2}
}
\providecommand{\href}[2]{#2}

\end{document}